\def\beq{\begin{equation}}
\def\eeq{\end{equation}}
\def\beqa{\begin{eqnarray}}
\def\eeqa{\end{eqnarray}}
\def\zero{{(0)}}
\def\one{{(1)}}
\def\two{{(2)}}
\def\three{{(3)}}
\def\four{{(4)}}
\def\LE{{\text{LE}}}
\newcommand\text[1]{\mbox{\scriptsize{#1}}}
\documentstyle[epsfig]{elsart}
\begin{document}
\begin{frontmatter}
\title{Non-Newtonian Poiseuille flow of a gas in a pipe}
{\bf Running title: Poiseuille flow}
\author{Mohamed Tij}
\address{D\'{e}partement de Physique, Universit\'{e} Moulay Isma\"{\i}l,\\
Mekn\`{e}s, Morocco}
\author{Andr\'es Santos}\thanks{Corresponding author: A. Santos, Tel.:
+34-924\,289\,540; fax: +34-924\,289\,651; E-mail address: andres@unex.es}
\address{Departamento de F\'{\i}sica, Universidad de Extremadura,\\
E--06071 Badajoz, Spain}

\date{\today}
\begin{abstract}
The Bhatnagar-Gross-Krook kinetic model of the Boltzmann equation is solved
for the steady  cylindrical Poiseuille flow fed by a
constant gravity field. The solution is obtained as a perturbation
expansion in powers of the field (through fourth order) and for a general
class of repulsive potentials.
The results, which are hardly sensitive to the interaction potential,
 suggest that the expansion is only asymptotic. A critical
comparison with the profiles predicted by the Navier-Stokes
 equations shows that the latter  fail over distances
comparable to the mean free path. In particular, while the Navier-Stokes
description predicts a monotonically decreasing
temperature as one moves apart from the cylinder axis, the
kinetic theory description shows that the temperature has a local minimum
 at the axis  and reaches a maximum value
 at a distance   of
the order of the mean free path. Within that distance, the radial heat flows
from the colder to the hotter points, in contrast to what is expected from 
the
Fourier law.
Furthermore, a longitudinal component of the heat flux exists in the absence
of gradients along the longitudinal direction.
Non-Newtonian effects, such as a non uniform
 hydrostatic pressure and  normal stress
differences, are also present.

\end{abstract}

\end{frontmatter}

\section{Introduction}
\label{sec1}
The steady flow in a long channel or in a long tube of circular section
under the action of a difference between the pressures imposed at the two
ends, usually known as Poiseuille flow or Hagen-Poiseuille flow, is a
typical textbook example in fluid dynamics \cite{textb}.
In the last few years, a number of authors
\cite{KMZ87,ELM94,AS92,TS94,MBG97,TTE97,TSS98,RC98,UG99,HM99} have analyzed
this problem with the channel geometry when the pressure difference is
replaced by a constant external field ${\bf g}$.
Kadanoff {\em et al.} \cite{KMZ87} have simulated this flow with the FHP
lattice gas automaton \cite{FHP86} to confirm the validity of a hydrodynamic
description for lattice gas automata.
For a dilute gas, Esposito {\em et al.} \cite{ELM94} have analyzed the
solution of the Boltzmann equation in the Navier-Stokes limit.
A generalized Navier-Stokes theory was seen to give a reasonable account of
a fluid composed of molecules that possess spin when compared with
molecular dynamics simulations \cite{TTE97}. Other studies
\cite{AS92,TS94,MBG97,TSS98,RC98,UG99,HM99}, on the other hand, have
focused on the breakdown of the continuum hydrodynamic predictions when the
strength of the external field is not asymptotically small. In Ref.\ 
\cite{AS92}, an
exact solution of the Bhatnagar-Gross-Krook (BGK) model kinetic equation was
found for a particular value of the field strength. The general solution
under the form of an expansion in powers of ${\bf g}$ was considered in
Ref.\ \cite{TS94}, where explicit expressions were derived to fifth order in
the field. More recently, the general solution corresponding to the
Boltzmann equation for Maxwell molecules has been derived to second order
\cite{TSS98} and  approximate solutions for hard spheres have been obtained
from a Burnett description \cite{UG99} and by means of  moment methods
\cite{RC98,HM99}. The theoretical predictions of Ref.\ \cite{TS94} have been
confirmed at a qualitative and semi-quantitative level by numerical
simulations of the Boltzmann equation \cite{MBG97,UG99} and by molecular
dynamics simulations \cite{RC98}.
The most surprising of those theoretical predictions is that the temperature
profile exhibits a bimodal shape, namely a local minimum at the middle of
the channel surrounded by two symmetric maxima at a distance of a
few mean free paths. In contrast, the continuum hydrodynamic equations
predict a temperature profile with a (flat) maximum at the middle. As a
consequence, the Fourier law is dramatically violated since in the slab
enclosed by the two maxima the tranverse component of the heat flux is
parallel (rather than anti-parallel) to the thermal gradient. Furthermore,
non-Newtonian properties, such as normal stress differences and
an effective shear viscosity depending on the hydrodynamic gradients, are
also present.

The goal of this paper is to carry out a kinetic theory description of the
Poiseuille flow driven by an external force when the gas is inside a pipe.
The reason is two-fold. First, the pipe geometry is much more realistic, and 
thus more worth studying, than
the channel one. Second, it is important to test whether the failure of the
continuum description to account for some of the qualitative features of the
Poiseuille flow in a channel  is not linked to  that particular
geometry and extends to the pipe case as well. In this context, it is
worthwhile noting that an exact solution of the Boltzmann equation (with
Maxwell molecules) for the {\em planar\/} Fourier flow, which is valid for 
arbitrary
values of the thermal gradient \cite{AMN79,SG95}, does {\em not\/} extend to
the {\em cylindrical\/} geometry \cite{AMN79}.
The results reported in this paper, on the other hand, confirm that the
structure of the hydrodynamic and flux profiles in the pipe problem is quite
similar to that of the channel problem. In particular, the temperature
exhibits a non-monotonic behavior as one moves apart from the pipe axis.
Nevertheless, the deviations from the continuum description are in general
quantitatively smaller than in the channel case.

The paper is organized as follows. The continuum hydrodynamic
description is worked out in Sec.\ \ref{sec2}, both for the
channel and cylindrical geometries. When the Navier-Stokes
constitutive equations are inserted into the exact balance
equations for mass, momentum and energy, a closed set of coupled
equations for the hydrodynamic fields (pressure, flow velocity and
temperature) is obtained. The spatial dependence of the transport
coefficients through the temperature is taken into
account. Given the nonlinear character of the set of equations,
its solution is expressed in powers of the external force. Section
\ref{sec3} is devoted to a summary of the results obtained from
the kinetic theory description in the case of the channel
Poiseuille flow and a critical comparison with the Navier-Stokes
predictions is carried out. The original part of the paper is
presented in Sec.\ \ref{sec4}, where the BGK kinetic equation is
solved for the pipe Poiseuille problem by means of a perturbation
expansion in powers of the force. Explicit expressions for the
successive contributions to the velocity distribution function
through fourth order in the force are derived. By velocity
integration, the profiles of the hydrodynamic fields and their
fluxes are then obtained. Since the content of Sec.\ \ref{sec4} is
rather technical, the discussion of the results is postponed to
Sec.\ \ref{sec5}, where only terms through third order are
considered. As in the planar case, the kinetic theory results
strongly differ from the continuum theory expectations, especially
in the case of the temperature profile. The breakdown of the
Fourier law is characterized by an apparent thermal conductivity
coefficient that only for large distances tends to the
Navier-Stokes coefficient. Analogously, an apparent shear
viscosity coefficient is introduced to monitor deviations from the
Newton law. Finally, the main conclusions of the paper are
presented in Sec.\ \ref{sec6}.

\section{Hydrodynamic description}
\label{sec2}
\subsection{Channel geometry}
\label{sec2A}
Let us first consider the Poiseuille flow in the channel geometry. A fluid
is enclosed between two infinite parallel
plates normal to the $y$ axis and located at $y=\pm H$, which are kept at
rest.
A constant external force per unit mass (e.g., gravity) ${\bf g}=g
\widehat{\bf z}$ is applied
along a direction $\widehat{\bf z}$ parallel to the plates.
We assume a laminar and incompressible regime, so  in the steady state 
the physical
quantities depend on  the coordinate $y$ only.
The balance equations for momentum and energy
become
\beq
\frac{\partial P_{yy}}{\partial y}=0,
\label{2.1b}
\eeq
\beq
\frac{\partial P_{yz}}{\partial y}=\rho  g,
\label{2.1a}
\eeq
\beq
P_{yz}\frac{\partial u_z}{\partial y}+
\frac{\partial q_{y}}{\partial y}=0,
\label{2.2}
\eeq
where $\rho$ is the mass density, ${\bf u}=u_z \widetilde{\bf z}$ is the 
flow velocity, ${\sf P}$ is
the pressure tensor and ${\bf q}$ is the heat flux.
In the Newtonian description these fluxes are related to the 
hydrodynamic gradients by the
 Navier-Stokes (NS) constitutive equations. In this problem they
 read
 \beq
P_{xx}=P_{yy}=P_{zz}=p,
\label{2.3a}
\eeq
\beq
P_{yz}=-\eta\frac{\partial u_{z}}{\partial y},
\label{2.3b}
\eeq
\beq
q_{y}=-\kappa\frac{\partial T}{\partial y},
\label{2.4a}
\eeq
\beq
q_z=0,
\label{2.4b}
\eeq
where $p=\frac{1}{3}\mbox{Tr}\, {\sf P}$ is the hydrostatic pressure, $T$ is
the temperature, and $\eta$ and $\kappa$ are the shear viscosity and the
thermal conductivity, respectively.
Combining Eqs.\ (\ref{2.1b})--(\ref{2.4a}), we get
\beq
\frac{\partial p}{\partial y}=0,
\label{2.5}
\eeq
\beq
\frac{\partial}{\partial y}\eta\frac{\partial u_z}{\partial y}=-\rho g,
\label{2.6}
\eeq
\beq
\frac{\partial}{\partial y}\kappa\frac{\partial T}{\partial y}=-
\eta\left(\frac{\partial u_z}{\partial y}\right)^2.
\label{2.7}
\eeq
Equation (\ref{2.6}) gives a parabolic-like velocity profile, that is
characteristic of the Poiseuille flow.
The temperature profile has, according to Eq.\ (\ref{2.7}), a quartic-like
shape.
Strictly speaking, these NS profiles are more complicated than just
polynomials
due to the temperature dependence of the transport coefficients.
Since the hydrodynamic profiles must be symmetric with respect to the plane
$y=0$, their odd derivatives must vanish at $y=0$. Thus, from Eqs.\
(\ref{2.6}) and (\ref{2.7}) we have
\beq
\left.\frac{\partial^2 u_z}{\partial y^2}\right|_{y=0}=-
\frac{\rho_0g}{\eta_0},
\label{2.8}
\eeq
\beq
\left.\frac{\partial^2 T}{\partial y^2}\right|_{y=0}=0,
\label{2.9}
\eeq
\beq
\left.\frac{\partial^4 T}{\partial y^4}\right|_{y=0}=-2
\frac{\rho_0^2g^2}{\eta_0\kappa_0},
\label{2.10}
\eeq
where  the subscript $0$  denotes quantities evaluated at $y=0$.
According to  Eqs.\ (\ref{2.8})--(\ref{2.10}), the NS equations
predict that  the flow velocity as well as the temperature have
a maximum at the middle layer $y=0$.
As we will see in the next Section, the kinetic theory description shows
that the temperature actually exhibits a local {\em minimum\/} at $y=0$,
since $\left.\partial^2T/\partial y^2\right|_{y=0}$ is a positive quantity
(of order $g^2$).

In order to get the hydrodynamic profiles from Eqs.\ (\ref{2.6}) and
(\ref{2.7}), one needs to know the density and temperature dependence of the
transport coefficients. To fix ideas, let us consider a {\em dilute\/} gas of
Maxwell molecules, in which case $p\propto \rho T$, $\eta\propto T$,
$\kappa\propto T$ \cite{CC70}. Consequently, from (\ref{2.5})--(\ref{2.7})
we get
\beq
p(y)=p_0,
\label{2.10.1}
\eeq
\beq
u_z(y)=u_0-\frac{\rho_0 g}{2\eta_0}\widetilde{y}^2,
\label{2.11}
\eeq
\beq
T(y)=T_0-\frac{\rho_0^2 g^2}{12\eta_0\kappa_0}\widetilde{y}^4,
\label{2.12}
\eeq
where $\widetilde{y}$ is an auxiliary space variable defined by
$d\widetilde{y}=[T_0/T(y)]dy$. From Eq.\ (\ref{2.12}) the relationship
between the true space variable $y$ and the scaled quantity $\widetilde{y}$
can be found as
\beq
y=\widetilde{y}\left(1-\frac{\rho_0^2
g^2}{60\eta_0\kappa_0T_0}\widetilde{y}^4\right).
\label{2.13}
\eeq
 Equations (\ref{2.3b})  and (\ref{2.4a}) then imply that
\beq
P_{yz}(y)=\rho_0 g\widetilde{y},
\label{2.13.1}
\eeq
\beq
q_y(y)=\frac{\rho_0^2 g^2}{4\eta_0}\widetilde{y}^3.
\label{2.13.2}
\eeq
The solution of the fifth-degree equation (\ref{2.13}), once inserted into
Eqs.\ (\ref{2.11}),  (\ref{2.12}),  (\ref{2.13.1}) and (\ref{2.13.2}),
yields the velocity, temperature and flux profiles predicted by the
NS equations in the case of Maxwell molecules. To fourth order in
the gravity field, the results are
\beq
u_z(y)=u_0-\frac{\rho_0
g}{2\eta_0}{y}^2\left(1+\frac{\rho_0^2g^2}{30\eta_0\kappa_0T_0}y^4\right)+
{\cal O}(g^5),
\label{2.11.1}
\eeq
\beq
T(y)=T_0-\frac{\rho_0^2 g^2}{12\eta_0\kappa_0}{y}^4
\left(1+\frac{\rho_0^2g^2}{15\eta_0\kappa_0T_0}y^4\right)+{\cal O}(g^6),
\label{2.12.1}
\eeq
\beq
P_{yz}(y)=\rho_0 g{y}\left(1+\frac{\rho_0^2
g^2}{60\eta_0\kappa_0T_0}{y}^4\right)+{\cal O}(g^5),
\label{2.11.3}
\eeq
\beq
q_{y}(y)=\frac{\rho_0^2 g^2}{3\eta_0}{y}^3\left(1+\frac{\rho_0^2
g^2}{20\eta_0\kappa_0T_0}{y}^4\right)+{\cal O}(g^6).
\label{2.11.4}
\eeq

It is
interesting to note that  $\widetilde{y}$ can be eliminated between Eqs.\
(\ref{2.11}) and (\ref{2.12}) to obtain the following
nonequilibrium ``equation of state'':
\beq
T=T_0-\frac{\eta_0}{3\kappa_0}(u_0-u_z)^2,
\label{2.14}
\eeq
which is independent of $g$.
By equation of state we mean in this context a relationship holding locally
among the hydrodynamic fields ($p$, $u_z$ and $T$) and that is independent
of gravity (at least up to a certain order). Since the pressure is uniform
in the NS description, it does not enter into Eq.\ (\ref{2.14}).
Interestingly enough, a quadratic dependence of the temperature on the flow
velocity also appears in the case of the steady planar Couette flow
\cite{MSG00}

\subsection{Cylindrical geometry}
\label{sec2B}
Now we assume that the fluid is inside a straight tube of uniform
circular section of radius $R$. Let the $z$ axis be parallel to the pipe
axis. As before, an external force per unit mass ${\bf g}=g\widehat{\bf
z}$ is applied to produce a flow field. In the (laminar) steady state all
the physically relevant quantities depend only on the distance $r\equiv
(x^2+y^2)^{1/2}$ from the axis. In the case of this cylindrical geometry,
the exact balance equations become
\beq
\frac{\partial}{\partial r}\left(r P_{rr}\right)=P_{\phi\phi},
\label{2.15a}
\eeq
\beq
r^{-1}\frac{\partial}{\partial r}\left(r P_{rz}\right)=\rho g,
\label{2.15b}
\eeq
\beq
P_{rz} r\frac{\partial
u_z}{\partial r}+\frac{\partial}{\partial r}\left(r q_r\right)=0.
\label{2.16}
\eeq
These equations constitute the cylindrical counterpart of Eqs.\
(\ref{2.1b})--(\ref{2.2}).
In general, the relationships between the cylindrical and Cartesian
components of a vector ${\bf A}$ and a tensor ${\sf B}$ are
\beq
\left(
\begin{array}{c}
A_r\\
A_\phi\\
A_z
\end{array}
\right)=
{\sf U}\cdot
\left(
\begin{array}{c}
A_x\\
A_y\\
A_z
\end{array}
\right),
\label{2.17}
\eeq
\beq
\left(
\begin{array}{ccc}
B_{rr}&B_{r\phi}&B_{rz}\\
B_{\phi r}&B_{\phi\phi}&B_{\phi z}\\
B_{zr}&B_{z\phi}&B_{zz}
\end{array}
\right)=
{\sf U}\cdot
\left(
\begin{array}{ccc}
B_{xx}&B_{xy}&B_{xz}\\
B_{yx}&B_{yy}&B_{y z}\\
B_{zx}&B_{zy}&B_{zz}
\end{array}
\right)
\cdot {\sf U}^\dagger,
\label{2.18}
\eeq
where
\beq
{\sf U}=
\left(
\begin{array}{ccc}
x/r&y/r&0\\
-y/r&x/r&0\\
0&0&1
\end{array}
\right)
\label{2.19}
\eeq
is a unitary matrix.

The NS constitutive equations yield
\beq
P_{rr}=P_{\phi\phi}=P_{zz}=p,
\label{2.20a}
\eeq
\beq
P_{rz}=-\eta\frac{\partial u_{z}}{\partial r},
\label{2.20b}
\eeq
\beq
q_{r}=-\kappa\frac{\partial T}{\partial r},
\label{2.21a}
\eeq
\beq
q_z=0.
\label{2.21b}
\eeq
The combination of Eqs.\ (\ref{2.15a})--(\ref{2.16}),
(\ref{2.20a})--(\ref{2.21a}) gives the following hydrodynamic equations:
\beq
\frac{\partial p}{\partial r}=0,
\label{2.22}
\eeq
\beq
r^{-1}\frac{\partial}{\partial r}\left(r\eta\frac{\partial u_z}{\partial
r}\right)=-\rho g,
\label{2.23}
\eeq
\beq
\frac{\partial}{\partial r}\left(r\kappa\frac{\partial T}{\partial
r}\right)=-\eta r \left(\frac{\partial u_z}{\partial
r}\right)^2.
\label{2.24}
\eeq
In contrast to what happens in the channel case, Eqs.\
(\ref{2.5})--(\ref{2.7}), it is not possible to obtain the explicit solution
to the hydrodynamic equations (\ref{2.22})--(\ref{2.24}), even with the help
of an auxiliary space variable.
On the other hand, the solution can be recursively found as a
series expansion in powers of $g$. To fourth order the result is
\beq
p(r)=p_0,
\label{2.24.1}
\eeq
\beq
u_z(r)=u_0-\frac{\rho_0
g}{4\eta_0}{r}^2\left[1+\frac{(4-3\alpha)\rho_0^2g^2}{576\eta_0\kappa_0T_0}
r^4\right]+{\cal
O}(g^5),
\label{2.25}
\eeq
\beq
T(r)=T_0-\frac{\rho_0^2 g^2}{64\eta_0\kappa_0}{r}^4
\left[1+\frac{(11-9\alpha)\rho_0^2g^2}{768\eta_0\kappa_0T_0}r^4\right]+{\cal
O}(g^6),
\label{2.26}
\eeq
where we have taken into account that $u_z$ and $T$ must be finite at $r=0$.
The
subscript 0 now denotes quantities evaluated at $r=0$.
Also, we have assumed that $\eta\propto T^{1-\alpha}$, $\kappa\propto
T^{1-\alpha}$, which corresponds to repulsive interaction potentials of
the form \cite{CC70} $\varphi(r)\propto r^{-\beta}$ with  
$\alpha=1/2-2/\beta$.
The cases $\alpha=0$ and $\alpha=\frac{1}{2}$ correspond to Maxwell
molecules ($\beta=4$) and hard spheres
($\beta\to\infty$), respectively.
The corresponding fluxes are
\beq
P_{rz}(r)=\frac{\rho_0 g}{2}{r}\left(1+\frac{\rho_0^2
g^2}{192\eta_0\kappa_0T_0}{r}^4\right)+{\cal O}(g^5),
\label{2.26.1}
\eeq
\beq
q_{r}(r)=\frac{\rho_0^2 g^2}{16\eta_0}{r}^3\left[1+\frac{(5-3\alpha)\rho_0^2
g^2}{384\eta_0\kappa_0T_0}{r}^4\right]+{\cal O}(g^6).
\label{2.26.2}
\eeq

For arbitrary $g$ the equation of state is not as simple as in the planar
case, Eq.\ (\ref{2.14}). Elimination of $r$ between Eqs.\ (\ref{2.25}) and
(\ref{2.26}) yields
\beq
T=T_0-\frac{\eta_0}{4\kappa_0}(u_0-u_z)^2-
\frac{(1-3\alpha)\eta_0^2}{576\kappa_0^2T_0}(u_0-u_z)^4+{\cal O}(g^6).
\label{2.27}
\eeq
\section{Kinetic theory description of the channel Poiseuille flow. A
summary}
\label{sec3}
The Poiseuille flow induced by an external force in the channel geometry has
been analyzed in the framework of kinetic theory
\cite{AS92,TS94,TSS98,RC98,UG99,HM99}, as well as by
numerical
simulations of the Boltzmann equation \cite{MBG97} and molecular dynamics
simulations \cite{RC98}. The emphasis in these papers, in contrast to
that of other works \cite{KMZ87,ELM94}, was put on highlighting the
limitations of the NS hydrodynamic description (see Sec.\
\ref{sec2A}) when the strength of the external field $g$ is not small enough.

In this Section we briefly summarize the main results derived in  Ref.\
\cite{TS94} from the BGK model of the Boltzmann equation. The BGK kinetic
equation reads \cite{C88}
\beq
\frac{\partial f}{\partial t}+{\bf v}\cdot \nabla
f+m^{-1}\frac{\partial}{\partial {\bf v}}\cdot ({\bf
F}f)=-\nu(f-f_{\text{LE}}),
\label{3.1}
\eeq
where $f({\bf r},{\bf v},t)$ is the one-particle velocity distribution
function, ${\bf F}$ is an external force, $\nu({\bf r},t)$ is an effective
collision frequency and
\beq
f_{\text{LE}}({\bf r},{\bf v},t)=n({\bf r},t)\left[\frac{m}{2\pi k_BT({\bf
r},t)}\right]^{3/2}\exp\left\{-\frac{m\left[{\bf v}-{\bf u}({\bf
r},t)\right]^2}{2k_BT({\bf r},t)}\right\}
\label{3.2}
\eeq
is the local equilibrium distribution function. Here, $m$ is the mass of a
particle, $k_B$ is the Boltzmann constant, $n({\bf r},t)$ is the local
number density, ${\bf u}({\bf r},t)$ is the local flow velocity and $T({\bf
r},t)$ is the local temperature. These hydrodynamic fields are defined as
velocity moments of $f$ by
\beq
n=\int d{\bf v}\, f,
\label{3.3}
\eeq
\beq
n{\bf u}=\int d{\bf v}\, {\bf v}f,
\label{3.4}
\eeq
\beq
nk_BT=\frac{m}{3}\int d{\bf v}\, V^2f,
\label{3.5}
\eeq
where in the last equation we have introduced the peculiar velocity ${\bf
V}={\bf v}-{\bf u}$. The fluxes of momentum and energy are characterized by
the pressure tensor
\beq
P_{ij}=m\int d{\bf v}\, V_iV_jf
\label{3.6}
\eeq
and the heat flux vector
\beq
{\bf q}=\frac{m}{2}\int d{\bf v}\, V^2{\bf V}f.
\label{3.7}
\eeq
The trace of the pressure tensor is $3p$, where $p=nk_BT$ is the (local)
hydrostatic pressure. The collision frequency $\nu$ is proportional to the
density and its dependence on the temperature changes in accordance with the
interaction potential considered. For instance, $\nu\propto nT^{1/2}$ for
hard spheres, while $\nu\propto n$ for Maxwell molecules.

For the steady Poiseuille flow in a channel, the BGK equation (\ref{3.1})
reduces to
\beq
\left(v_y\frac{\partial}{\partial y}+g\frac{\partial}{\partial
v_z}\right)f=-\nu(f-f_{\text{LE}}),
\label{3.8}
\eeq
which must be complemented with the appropriate boundary conditions at
$y=\pm H$. On the other hand, we assume that the separation between the
plates is large enough to allow for the existence of a {\em bulk\/} region
$-H+\delta<y<H-\delta$, where $\delta$ is the width of the boundary layers
and comprises a few mean free paths. Inside the bulk region the solution to
Eq.\ (\ref{3.8}) is expected to be rather insensitive to the details of
the boundary conditions and depend on $y$ through a functional dependence on
the hydrodynamic fields. Such a solution was obtained in Ref.\ \cite{TS94}
(for Maxwell molecules) by means of a perturbation expansion in powers of
$g$. Here we quote the hydrodynamic fields through third order:
\beq
p(y)=p_0\left[1+\zeta_p\left(\frac{mg}{k_BT_0}\right)^2y^2\right]+
{\cal O}(g^4),
\label{3.9}
\eeq
\beqa
u_z(y)&=&u_0-\frac{\rho_0
g}{2\eta_0}{y}^2\left[1+\frac{\rho_0^2g^2}{30\eta_0\kappa_0T_0}y^4
\right.\nonumber\\
&&\left.
+\zeta_u\left(\frac{mg}{k_BT_0}\right)^2y^2
+\zeta_u'\frac{\rho_0\eta_0^2g^2}{p_0^3}
\right]+{\cal O}(g^5),
\label{3.10}
\eeqa
\beq
T(y)=T_0\left[1-\frac{\rho_0^2 g^2}{12\eta_0\kappa_0T_0}{y}^4+
\zeta_T\left(\frac{mg}{k_BT_0}\right)^2y^2\right]
+{\cal O}(g^4),
\label{3.11}
\eeq
where $\zeta_p=\frac{6}{5}=1.2$, $\zeta_u=\frac{152}{25}=6.08$,
$\zeta_u'=\frac{5474}{25}=218.96$, and $\zeta_T=\frac{19}{25}=0.76$.
The BGK predictions (\ref{3.9})--(\ref{3.11}) have been confirmed by an
exact  solution of the Boltzmann equation for Maxwell molecules
\cite{TSS98} ($\zeta_p=\frac{6}{5}$, $\zeta_T\simeq 1.0153$), as well as by
approximate solutions of the Boltzmann equation for hard spheres by a
13-moment method \cite{RC98,HM99} ($\zeta_p=\frac{6}{5}$,
$\zeta_T=\frac{14}{25}=0.56$) and a 19-moment method \cite{HM99}
($\zeta_p\simeq 1.214$, $\zeta_T\simeq 0.99$).
Comparison with the NS predictions, Eqs.\ (\ref{2.10.1}),
(\ref{2.11.1}) and (\ref{2.12.1}), shows that the latter already fail to
second order ($\zeta_p^{\text{NS}}=\zeta_T^{\text{NS}}=0$)
and to third order
($\zeta_u^{\text{NS}}=\zeta_u'^{\text{NS}}=0$) in $g$.
According to Eqs.\ (\ref{3.9})--(\ref{3.11}), the pressure increases
 parabolically from the midpoint ($\zeta_p>0$) rather than being uniform,
 the
velocity profile has an enhanced quadratic coefficient ($\zeta_u'>0$) plus a
new quartic term ($\zeta_u>0$), and the temperature has a positive quadratic
term ($\zeta_T>0$). The latter is responsible for the fact that
$\left.\partial^2 T/\partial y^2\right|_{y=0}>0$, in contrast to the
NS prediction (\ref{2.9}), so  the temperature presents a local
{\em minimum\/} at $y=0$ rather than a maximum.
This minimum is surrounded by two maxima located at
$y=y_{\text{max}}=\pm\sqrt{6\zeta_T}\ell_0$, where $\ell_0\equiv
(\eta_0\kappa_0 T_0)^{1/2}/p_0$ is a reference mean free path \cite{HM99}.
 The relative difference
between the maxima and the minimum is
$(T_{\text{max}}-T_0)/T_0=3\zeta_T^2(\ell_0/h_0)^2$, where $h_0\equiv
(k_BT_0/m)/g$ is the so-called scale height \cite{textbooks}, i.e. the
characteristic distance associated with the external (gravity) field.
This surprising bimodal form of the temperature profile is an effect going
beyond the Burnett description \cite{UG99} and has been  confirmed by
Monte Carlo simulations of the Boltzmann equation for hard spheres
\cite{MBG97,UG99}.
The NS equation of state (\ref{2.14}) is now augmented by an extra
term:
\beq
T=T_0-\frac{\eta_0}{3\kappa_0}(u_0-u_z)^2+
\frac{\zeta_T}{\zeta_p}\frac{T_0}{p_0}(p-p_0)+{\cal O}(g^4).
\label{3.12}
\eeq

In addition to the hydrodynamic profiles, the momentum and heat fluxes are
obtained from the kinetic theory description. In the case of the BGK model,
the results are \cite{TS94}:
\beq
P_{xx}(y)=p_0\left[1-\frac{22}{25}\frac{\rho_0\eta_0^2g^2}{p_0^3}+\frac{4}{5}
\left(\frac{mg}{k_BT_0}\right)^2y^2\right]+
{\cal O}(g^4),
\label{3.13}
\eeq
\beq
P_{yy}(y)=p_0\left(1-\frac{306}{25}\frac{\rho_0\eta_0^2g^2}{p_0^3}\right)+
{\cal O}(g^4),
\label{3.14}
\eeq
\beq
P_{zz}(y)=p_0\left[1+\frac{328}{25}\frac{\rho_0\eta_0^2g^2}{p_0^3}
+\frac{14}{5}
\left(\frac{mg}{k_BT_0}\right)^2y^2\right]+
{\cal O}(g^4),
\label{3.15}
\eeq
\beq
P_{yz}(y)=\rho_0gy\left[1+\frac{\rho_0^2
g^2}{60\eta_0\kappa_0T_0}{y}^4+\frac{11}{75}
\left(\frac{mg}{k_BT_0}\right)^2y^2\right]+
{\cal O}(g^5),
\label{3.16}
\eeq
\beq
q_y(y)=\frac{\rho_0^2g^2}{3\eta_0}y^3+{\cal O}(g^4),
\label{3.17}
\eeq
\beqa
q_z(y)&=&-\frac{2m g
\kappa_0}{5k_B}\left[1-\frac{21162\rho_0\eta_0^2 g^2}{25p_0^3}
-\frac{159}{5}\left(\frac{mg}{k_BT_0}\right)^2y^2
\right.\nonumber\\
&&\left.-\frac{29\rho_0^2
g^2}{12\eta_0\kappa_0T_0}{y}^4
\right]
+{\cal O}(g^5),
\label{3.18}
\eeqa
As expected, the fluxes differ from the NS results, Eqs.\
(\ref{2.3a}), (\ref{2.4b}), (\ref{2.11.3}) and (\ref{2.11.4}).
The main deviations of the hydrodynamic and flux profiles from the
NS predictions occur for distances on the scale of the mean free
path, i.e. in the regime where a hydrodynamic description is not expected to
hold. For instance, the extra terms appearing in Eq.\ (\ref{3.10}) are,
relative to the $g^2$-term of Eq.\ (\ref{2.11.1}), of orders
$(y/\ell_0)^{-2}$ and $(y/\ell_0)^{-4}$. In addition, the ratio between the
component of the heat flux parallel to the flow direction ($q_z$) and the
component parallel to the thermal gradient ($q_y$) is of order
$(y/\ell_0)^{-3}(h_0/\ell_0)$.
Thus, the NS description applies in the regime $(y/\ell_0)\gg
(h_0/\ell_0)^{1/3}\gg 1$. On the other hand, the kinetic theory description,
while limited here to weak fields, i.e. $(h_0/\ell_0)\gg 1$, is still valid
for $y\sim \ell_0$.

\section{Kinetic theory description of the pipe Poiseuille flow}
\label{sec4}
Now we are going to analyze the solution of the BGK equation for the
steady Poiseuille flow problem in a cylindrical geometry. In that case, Eq.\
(\ref{3.1}) becomes
\beq
\left(v_r\frac{\partial}{\partial
r}+\frac{v_\phi}{r}\frac{\partial}{\partial \phi}+g\frac{\partial}{\partial
v_z}\right)f=-\nu (f-f_\LE).
\label{4.1}
\eeq
Note that the derivative $\partial/\partial\phi$ is understood at constant
$(v_x,v_y)$, {\em not\/} at constant $(v_r,v_\phi)$. In fact, $\partial
v_r/\partial\phi=v_\phi$ and $\partial v_\phi/\partial\phi=-v_r$, and so 
\beq
\frac{\partial f}{\partial\phi}=v_\phi \frac{\partial f}{\partial v_r}-v_r
\frac{\partial f}{\partial v_\phi}.
\label{4.2}
\eeq
The exact conservation equations (\ref{2.15a})--(\ref{2.16}) can be
reobtained from Eq.\ (\ref{4.1}) by multiplying both sides by $v_r$, $v_z$
and $V^2$, respectively, and integrating over the velocity.

As in Sec.\ \ref{sec2}, we denote by a subscript 0 those quantities
evaluated at the axis of the pipe ($r=0$). Thus, $v_0\equiv
(k_BT_0/m)^{1/2}$ is a thermal velocity, $\lambda_0\equiv v_0/\nu_0$ is a
mean free path and $h_0\equiv v_0^2/g$ is a characteristic length associated
with gravity (scale height). Since in the BGK model $\eta=p/\nu$ and
$\kappa=5k_B\eta/2m$, one has $\ell_0=\sqrt{\frac{5}{2}}\lambda_0$, where
$\ell_0$ was introduced in the previous Section.
Without loss of generality, we will assume a reference frame stationary with
the flow at $r=0$, so  $u_0=0$. We next introduce dimensionless
quantities as
\beq
{\bf v}^*=v_0^{-1}{\bf v},\quad {\bf r}^*=\lambda_0^{-1}{\bf r}, \quad
f^*=n_0^{-1}v_0^3 f,
\label{4.3a}
\eeq
\beq
p^*=p_0^{-1}p,\quad {\bf u}^*=v_0^{-1}{\bf u}, \quad T^*=T_0^{-1}T,
\label{4.3b}
\eeq
\beq
\nu^*=\nu_0^{-1}\nu,\quad g^*=(v_0\nu_0)^{-1}g=\lambda_0/h_0.
\label{4.3c}
\eeq
In order to simplify the notation, the asterisks will be dropped henceforth,
so all the quantities will be understood to be expressed in reduced units,
unless stated otherwise.

The objective now is to find the solution to Eq.\ (\ref{4.1}) as an
expansion in powers of $g$:
\beq
f=f^\zero+f^\one g+f^\two g^2+f^\three g^3+\cdots,
\label{4.4}
\eeq
where $f^\zero$ is the equilibrium distribution function normalized to
$p^\zero=1$, $T^\zero=1$. Similar expansions hold for the hydrodynamic
fields:
\beq
p=1+p^\two g^2+p^{(4)} g^4+\cdots,
\label{4.5}
\eeq
\beq
u_z=u^\one g+u^\three g^3+\cdots,
\label{4.6}
\eeq
\beq
T=1+T^\two g^2+T^{(4)} g^4+\cdots,
\label{4.7}
\eeq
where we have taken into account that, because of the symmetry of the
problem, $p$ and $T$ are even functions of $g$, while $u_z$ is an odd
function. Insertion of Eq.\ (\ref{4.4}) into Eq.\ (\ref{4.1}) yields
\beq
\left(1+{\cal A}\right)f^{(n)}=f_{\text{LE}}^{(n)}-{\cal D}f^{(n-1)}-
\sum_{m=1}^{n-2}\nu^{(n-m)}\left(f^{(m)}-f_{\text{LE}}^{(m)}\right),
\label{4.8}
\eeq
where the operators ${\cal A}$ and ${\cal D}$ are defined as
\beq
{\cal A}\equiv v_r\frac{\partial}{\partial r}+\frac{v_\phi}{r}\left(
v_\phi\frac{\partial}{\partial v_r}-v_r\frac{\partial}{\partial
v_\phi}\right),
\label{4.9}
\eeq
\beq
{\cal D}\equiv\frac{\partial}{\partial v_z}.
\label{4.10}
\eeq
The formal solution to Eq.\ (\ref{4.8}) is
\beq
f^{(n)}=\sum_{k=0}^\infty (-{\cal A})^k\left[f_\LE^{(n)}-{\cal D} f^{(n-1)}-
\sum_{m=1}^{n-2}\nu^{(n-m)}\left(f^{(m)}-f_{\text{LE}}^{(m)}\right)\right].
\label{4.11}
\eeq
This solution is not complete because $f^{(n)}$ appears implicitly on the
right side through the dependence of $f^{(n)}_\LE$ on $p^{(n)}$, $u^{(n)}$
and $T^{(n)}$. If the space dependence of these quantities were known, Eq.\
(\ref{4.11}) would give us $f^{(n)}$, provided that the previous
contributions $\{f^{(m)}, m\leq n-1\}$ are known. In order to get a closed
set of equations for $p^{(n)}$, $u^{(n)}$ and $T^{(n)}$, we must apply the
consistency conditions
\beq
\int d{\bf v}\, \left(f^{(n)}-f^{(n)}_\LE\right)=0,
\label{4.12}
\eeq
\beq
\int d{\bf v}\, v_z\left(f^{(n)}-f^{(n)}_\LE\right)=0,
\label{4.13}
\eeq
\beq
\int d{\bf v}\, v^2\left(f^{(n)}-f^{(n)}_\LE\right)=0.
\label{4.14}
\eeq
\subsection{First-order results}
\label{sec4A}
In this case, $f^\one_\LE=-u^\one{\cal D}f^\zero$, and so Eq.\ (\ref{4.11})
yields
\beq
f^\one=f^\one_\LE-{\cal D}\left[f^\zero+\sum_{k=1}^\infty (-{\cal A})^k
u^\one f^\zero\right],
\label{4.15}
\eeq
where we have taken into account that the operators ${\cal A}$ and ${\cal
D}$ commute and that ${\cal A}^k f^\zero=0$ for $k\geq 1$.
The conditions (\ref{4.12}) and (\ref{4.14}) are automatically satisfied. As
for condition (\ref{4.13}), it implies that
\beq
\int d{\bf v}\, \sum_{k=1}^\infty (-{\cal A})^k
u^\one f^\zero=-1.
\label{4.16}
\eeq
The simplest solution to Eq.\ (\ref{4.16}) is expected to be of the form
\beq
u^\one(r)=u_{12}r^2.
\label{4.17}
\eeq
To confirm this, let us express the operator ${\cal A}$ in Cartesian
coordinates, i.e. ${\cal A}=v_x\partial/\partial x+v_y\partial/\partial y$.
Consequently, only the terms with $k\leq 2$ contribute in Eqs.\ (\ref{4.15})
and (\ref{4.16}). Insertion of (\ref{4.17}) into (\ref{4.16}) then gives
\beq
u_{12}=-\frac{1}{4}.
\label{4.17.1}
\eeq
The explicit expression for $f^\one$ is simply
\beq
f^\one=v_z\left[1-\frac{1}{2}(v_r^2+v_\phi^2)+\frac{1}{2} r
v_r-\frac{1}{4}r^2\right]f^\zero.
\label{4.18}
\eeq
The nonzero components of the pressure tensor and the heat flux are, to
first order,
\beq
P_{rz}^\one=\int d{\bf v}\, v_r v_z f^\one
=\frac{1}{2} r,
\label{4.19}
\eeq
\beq
q_{z}^\one=\int d{\bf v}\, \left[\frac{1}{2}v^2 v_z
f^\one-u^\one\left(v_z^2+\frac{1}{2}v^2\right)f^\zero\right]
=-1.
\label{4.20}
\eeq
\subsection{Second-order results}
\label{sec4B}
The second-order contribution to the distribution function is
\beq
f^\two=f^\two_\LE-{\cal D}f^\one+\sum_{k=1}^\infty (-{\cal A})^k
\left(f^\two_\LE-{\cal D} f^\one\right),
\label{4.21}
\eeq
where
\beq
f^\two_\LE=\left[\frac{1}{2}{u^\one}^2(v_z^2-1)+
p^\two+\frac{1}{2}T^\two(v^2-5)\right]f^\zero.
\label{4.22}
\eeq
Equation (\ref{4.13}) is identically satisfied, since both $f^\two_\LE$ and
${\cal D}f^\one$ are even functions of $v_z$. Conditions
(\ref{4.12}) and (\ref{4.14}) become, respectively,
\beq
\int d{\bf v}\, \sum_{k=1}^\infty (-{\cal A})^k
 f^\two_\LE=0,
\label{4.23}
\eeq
\beqa
\int d{\bf v}\, v^2 \sum_{k=1}^\infty (-{\cal A})^k
 f^\two_\LE&=&-2\int d{\bf v}\, v_z \left(1-{\cal A}+{\cal A}^2\right)f^\one
 \nonumber\\
 &=&4+\frac{1}{2}r^2.
\label{4.24}
\eeqa
Next, by looking for a solution with a spatial dependence similar to that of
the planar case, Eqs.\ (\ref{3.9})--(\ref{3.11}), we write
\beq
p^\two(r)=p_{22}r^2,
\label{4.25}
\eeq
\beq
T^\two(r)=T_{22}r^2+T_{24}r^4.
\label{4.26}
\eeq
Consequently, only the terms with $k\leq 4$ contribute in Eqs.\ (\ref{4.23})
and (\ref{4.24}). More explicitly,
\beqa
\sum_{k=1}^\infty (-{\cal A})^k f^\two_\LE&=&
\left\{\left[p_{22}+\frac{T_{22}}{2}(v^2-5)\right]
\left[-2rv_r+2(v_r^2+v_\phi^2)\right]\right.\nonumber\\
&&
+\left[\frac{1}{32}(v_z^2-1)+\frac{T_{24}}{2}(v^2-5)\right]
\left[-4r^3v_r+4r^2(3v_r^2+v_\phi^2)\right.\nonumber\\
&&\left.\left.
-24rv_r(v_r^2+v_\phi^2)+24(v_r^2+v_\phi^2)^2\right]\right\}f^\zero.
\label{4.27}
\eeqa
Insertion into Eqs.\ (\ref{4.23}) and (\ref{4.24}) yields
\beq
p_{22}+48T_{24}=0,
\label{4.28}
\eeq
\beq
20p_{22}+20T_{22}+2688T_{24}+12+(80T_{24}+1)r^2=4+\frac{1}{2}r^2,
\label{4.29}
\eeq
respectively. The solution is
\beq
p_{22}=\frac{3}{10},
\label{4.29.1}
\eeq
\beq
T_{22}=\frac{7}{50},\quad
T_{24}=-\frac{1}{160}.
\label{4.29.2}
\eeq
 The explicit expression for $f^\two$ is then
\beqa
f^\two&=&\sum_{k=0}^4(-{\cal A})^k f^\two_\LE-{\cal D}\sum_{k=0}^2(-{\cal
A})^k f^\one\nonumber\\
&=&\left\{(v_z^2-1)\left[1-\frac{3}{2}(v_r^2+v_\phi^2)+rv_r-
\frac{1}{4}r^2\right]\right.
\nonumber\\
&&
+\frac{1}{10}\left[3+\frac{7}{10}(v^2-5)\right]\left[2(v_r^2+v_\phi^2)
-2rv_r+r^2\right]
\nonumber\\
&&
+\frac{1}{32}\left[v_z^2-1-\frac{1}{10}(v^2-5)\right]
\left[24(v_r^2+v_\phi^2)^2-24rv_r(v_r^2+v_\phi^2)
\right.
\nonumber\\
&&\left.\left.
+4r^2(3v_r^2+v_\phi^2)
-4r^3v_r+r^4\right]\right\}f^\zero.
\label{4.30}
\eeqa
The second-order contributions to the pressure tensor are
\beq
P_{rr}^\two=\int d{\bf v}\, v_r^2 f^\two
=-\frac{92}{25}+\frac{1}{20} r^2,
\label{4.31}
\eeq
\beq
P_{\phi\phi}^\two=\int d{\bf v}\, v_\phi^2 f^\two
=-\frac{92}{25}+\frac{3}{20} r^2,
\label{4.32}
\eeq
\beq
P_{zz}^\two=\int d{\bf v}\, v_z^2 f^\two-{u^\one}^2
=\frac{184}{25}+\frac{7}{10} r^2.
\label{4.33}
\eeq
Analogously,
\beq
q_{r}^\two=\frac{1}{2}\int d{\bf v}\, v^2 v_r f^\two-u^\one P_{rz}^\one
=\frac{1}{16} r^3.
\label{4.33.1}
\eeq
\subsection{Third- and fourth-order results}
\label{sec4C}
For $n=3$, Eq.\ (\ref{4.11}) reduces to
\beq
f^\three=
\sum_{k=0}^\infty (-{\cal A})^k
\left[f^\three_\LE-{\cal D}
f^\two-\nu^\two\left(f^\one-f^\one_\LE\right)\right],
\label{4.34}
\eeq
where
\beq
f^\three_\LE=v_z\left[u^\one f^\two_\LE+\left(u^\three-u^\one
T^\two-\frac{1}{3}{u^\one}^3v_z^2\right)f^\zero\right].
\label{4.35}
\eeq
Up to now, we have not needed to fix the temperature dependence of the
collision frequency. This implies that to second order in the field the
results are {\em universal}, i.e. independent of the interaction potential.
On the other hand, the results of higher order are sensitive to the
potential. For the sake of concreteness, we now consider
repulsive interaction potentials of the form $\varphi(r)\propto r^{-\beta}$, 
for
which the collision frequency  is \cite{C88} $\nu\propto p
T^{-(1-\alpha)}$ with $\alpha=1/2-2/\beta$. In particular, the case
$\alpha=0$ corresponds to Maxwell molecules ($\beta=4$), while the case
$\alpha=\frac{1}{2}$ refers to hard spheres ($\beta\to\infty$). For this
class of potentials, $\nu^\two=p^\two-(1-\alpha)T^\two$.

Conditions (\ref{4.12}) and (\ref{4.14}) with $n=3$ are
identically satisfied because of symmetry. The structure of
$u^\one$, $p^\two$ and $T^\two$ suggests that $u^\three$ has a
spatial dependence of the form 
\beq
u^\three(r)=u_{32}r^2+u_{34}r^4+u_{36}r^6, 
\label{4.36} 
\eeq 
so
 only the terms with $k\leq 6$ contribute in Eq.\
(\ref{4.34}). Insertion into Eq.\ (\ref{4.13}) yields 
\beqa &&
34560u_{36}+192u_{34}+4u_{32}+\frac{2(1367-206\alpha)}{25}
+\left(1728u_{36}+16u_{34}\right. \nonumber\\ 
&&
\left.+\frac{149-36\alpha}{50}\right)r^2+
\left(36u_{36}+\frac{4-3\alpha}{160}\right)r^4=0, 
\label{4.37}
\eeqa 
whose solution is 
\beq
u_{32}=-\frac{4(100-\alpha)}{25},\quad
u_{34}=-\frac{89+9\alpha}{800},\quad
u_{36}=-\frac{4-3\alpha}{5760}. 
\label{4.38} 
\eeq 
Once $u^\three$
is determined, Eq.\ (\ref{4.34}) gives the explicit form of
$f^\three$. From it we can easily get \beq P_{rz}^\three=\int
d{\bf v}\, v_r v_zf^\three=\frac{1}{25}r^3+\frac{1}{960}r^5,
\label{4.41} \eeq \beqa q_{z}^\three&=&\frac{1}{2}\int d{\bf v}\,
v^2 v_zf^\three-\frac{5}{2}u^\three-u^\one\left(\frac{3}{2}
p^\two+\frac{1}{2}{u^\one}^2+P_{zz}^\two\right)\nonumber\\
&=&\frac{4(1358-23\alpha)}{25}+\frac{209-3\alpha}{50}r^2 +
\frac{15-\alpha}{160}r^4. \label{4.42} \eeqa

Proceeding in a similar way, higher order terms can be evaluated, but the
algebra becomes progressively more cumbersome. Here we only quote the main
results to fourth order in $g$. Equation (\ref{4.11}) gives
\beq
f^\four=
\sum_{k=0}^\infty (-{\cal A})^k
\left[f^\four_\LE-{\cal D}
f^\three-\nu^\two\left(f^\two-f^\two_\LE\right)\right].
\label{4.43}
\eeq
By assuming that
\beq
p^\four(r)=p_{42}r^2+p_{44}r^4+p_{46}r^6,
\label{4.44}
\eeq
\beq
T^\four(r)=T_{42}r^2+T_{44}r^4+T_{46}r^6+T_{48}r^8,
\label{4.45}
\eeq
it turns out that only the terms with $k\leq 8$ contribute in Eq.\
(\ref{4.43}). By symmetry, condition (\ref{4.13}) is identically satisfied.
On the other hand, conditions (\ref{4.12}) and (\ref{4.14}) give,
respectively,
\beqa
&&
4p_{42}+192p_{44}+34560p_{46}+192T_{44}+69120T_{46}+46448640T_{48}
\nonumber\\
&&
+\frac{8(243511-103801\alpha)}{625}+\left[\frac{14(95-56\alpha)}{25}+
16p_{44}+1728p_{46}+1728 T_{46}\right.
\nonumber\\
&&\left.+1105920T_{48}\right]r^2
+\left[\frac{39(1-\alpha)}{200}+
36p_{46}+6912T_{48}\right]r^4=0,
\label{4.46}
\eeqa
\beqa
&&
\frac{4(448099-1773928\alpha)}{625}+20T_{42}+1728T_{44}+546048T_{46}+
403881984T_{48}
\nonumber\\
&&
-\left[\frac{134783-59955\alpha}{250}+
80T_{44}+15552 T_{46}
+8736768T_{48}\right]r^2
\nonumber\\
&&
+\left[\frac{1345-642\alpha}{400}+
180T_{46}+62208T_{48}\right]r^4
\nonumber\\
&&
+\left[\frac{11-9\alpha}{960}+
320T_{48}\right]r^6=0,
\label{4.47}
\eeqa
where in Eq.\ (\ref{4.47}) we have eliminated $p_{42}$--$p_{46}$ in favor of
$T_{42}$--$T_{48}$. The solution to Eqs.\ (\ref{4.46}) and (\ref{4.47}) is
\beq
p_{42}=-\frac{218083-11035\alpha}{1250},\quad
p_{44}=-\frac{653-176\alpha}{2000},\quad
p_{46}=\frac{7-\alpha}{4800},
\label{4.48}
\eeq
\beqa
&&T_{42}=-\frac{2501129-38495\alpha}{6250},\quad
T_{44}=-\frac{32057-663\alpha}{20000},
\nonumber\\
&&T_{46}=-\frac{454+87\alpha}{72000},\quad
T_{48}=-\frac{11-9\alpha}{307200}.
\label{4.54}
\eeqa

From Eq.\ (\ref{4.43}) we can now evaluate the fourth-order terms in the
pressure tensor and the heat flux. The results are
\beqa
P_{rr}^\four&=&\int d{\bf v}\,
v_r^2f^\four=\frac{4(5087846-175355\alpha)}{3125}
\nonumber\\
&&
-\frac{106029-3653\alpha}{2500}r^2
-\frac{287-18\alpha}{6000}r^4
+\frac{3-\alpha}{19200}r^6,
\label{4.55}
\eeqa
\beqa
P_{\phi\phi}^\four&=&\int d{\bf v}\,
v_\phi^2f^\four=\frac{4(5087846-175355\alpha)}{3125}
\nonumber\\
&&
-\frac{3(106029-3653\alpha)}{2500}r^2
-\frac{287-18\alpha}{1200}r^4
+\frac{7(3-\alpha)}{19200}r^6,
\label{4.56}
\eeqa
\beqa
P_{zz}^\four&=&\int d{\bf v}\,
v_z^2f^\four-2u^\one u^\three-{u^\one}^2\left(p^\two-T^\two\right)
\nonumber\\
&=&-\frac{8(5087846-175355\alpha)}{3125}
-\frac{442191-25799\alpha}{1250}r^2
\nonumber\\
&&
-\frac{1385-492\alpha}{2000}r^4
+\frac{15-\alpha}{4800}r^6,
\label{4.57}
\eeqa
\beqa
q_{r}^\four&=&\frac{1}{2}\int d{\bf v}\,
v^2v_rf^\four-u^\one P_{rz}^\three-u^\three
P_{rz}^\one
\nonumber\\
&=&\frac{100-\alpha}{25}r^3
+\frac{97+9\alpha}{2400}r^5
+\frac{5-3\alpha}{15360}r^7.
\label{4.58}
\eeqa

It can be checked that the results for $p(r)$, $T(r)$, $u_z(r)$,
$P_{rr}(r)$, $P_{\phi\phi}(r)$, $P_{r\phi}(r)$ and $q_{r}(r)$ we have
derived are indeed consistent with the balance equations
(\ref{2.15a})--(\ref{2.16}). In fact, Eq.\ (\ref{2.15b}) allows us to
obtain $P_{rz}^{(5)}(r)$ with the result
\beqa
P_{rz}^{(5)}&=&\frac{3(235119+2780\alpha)}{12500}r^3
+\frac{25079+1097\alpha}{120000}r^5
\nonumber\\
&&
+\frac{71+9\alpha}{72000}r^7
+\frac{23-9\alpha}{3072000}r^9.
\label{4.59}
\eeqa
\section{Discussion}
\label{sec5}
When Eqs.\ (\ref{4.17}), (\ref{4.17.1}), (\ref{4.25}), (\ref{4.26}),
(\ref{4.29.1}), (\ref{4.29.2}), (\ref{4.36}), (\ref{4.38}), (\ref{4.44}),
(\ref{4.45}), (\ref{4.48}) and (\ref{4.54}) are inserted into Eqs.\
(\ref{4.5})--(\ref{4.7}), one gets the hydrodynamic profiles predicted by
the BGK kinetic model through fourth order in the field. Comparison with the
NS predictions, Eqs.\ (\ref{2.24.1})--(\ref{2.26}), indicates that
the latter, while  providing the correct values of $u_{12}$, $u_{36}$,
$T_{24}$ and $T_{48}$, do not capture the pressure variation
($p_{22}^{\text{NS}}=p_{42}^{\text{NS}}=p_{44}^{\text{NS}}=p_{46}^{\text{NS}}=0$)
or the lower-degree terms of the velocity
($u_{32}^{\text{NS}}=u_{34}^{\text{NS}}=0$) and temperature
($T_{22}^{\text{NS}}=T_{42}^{\text{NS}}=T_{44}^{\text{NS}}=T_{46}^{\text{NS}}=0$)
profiles.

The results derived in the previous Section strongly support the conjecture
that the expansion in powers of $g$ is asymptotic rather than convergent
\cite{TS94}. For instance $T=1+0.14g^2-4.0\times 10^2 g^4+\cdots$ and
$T=1-0.41g^2-1.1\times 10^4 g^4+\cdots$ at $r=1$ and $r=5$, respectively.
Thus, the expansion is only useful if $g$ is small enough to keep the first
few terms only. As we did in the planar case, Sec.\ \ref{sec3}, we now give
the hydrodynamic profiles through third order in real units:
\beq
p(r)=p_0\left[1+\zeta_p\left(\frac{mg}{k_BT_0}\right)^2r^2\right]+
{\cal O}(g^4),
\label{5.1}
\eeq
\beqa
u_z(r)&=&u_0-\frac{\rho_0
g}{4\eta_0}{r}^2\left[1+\frac{(4-3\alpha)\rho_0^2g^2}{576\eta_0\kappa_0T_0}r^4
\right.\nonumber\\
&&\left.
+\zeta_u\left(\frac{mg}{k_BT_0}\right)^2r^2
+\zeta_u'\frac{\rho_0\eta_0^2g^2}{p_0^3}
\right]+{\cal O}(g^5),
\label{5.2}
\eeqa
\beq
T(r)=T_0\left[1-\frac{\rho_0^2 g^2}{64\eta_0\kappa_0T_0}{r}^4+
\zeta_T\left(\frac{mg}{k_BT_0}\right)^2r^2\right]
+{\cal O}(g^4),
\label{5.3}
\eeq
where $\zeta_p=\frac{3}{10}$, $\zeta_u=(89+9\alpha)/200$,
$\zeta_u'=\frac{16}{25}(100-\alpha)$ and $\zeta_T=\frac{7}{50}$.
The structure of these profiles is similar to that of the planar case, Eqs.\
(\ref{3.9})--(\ref{3.11}).
In particular, the temperature presents a local
{minimum} at $r=0$  and a maximum  at
$r=\sqrt{32\zeta_T}\ell_0\equiv r_{\text{max}}$, where $\ell_0\equiv
(\eta_0\kappa_0 T_0)^{1/2}/p_0$.
 The relative difference
between the maximum and the minimum is
$(T_{\text{max}}-T_0)/T_0=16\zeta_T^2(\ell_0/h_0)^2$, where $h_0\equiv
(k_BT_0/m)/g$.
The distance from the center of the points where the temperature reaches the
value $T_{\text{max}}$ in the pipe flow ($r_{\text{max}}\simeq 2.12 \ell_0$)
is very close to
that of the channel flow ($y_{\text{max}}\simeq \pm 2.14 \ell_0$).
On the other hand, the
effect is considerably smaller in the former case
[$(T_{\text{max}}-T_0)/T_0\simeq 0.31(\ell_0/h_0)^2$]
than in the latter
[$(T_{\text{max}}-T_0)/T_0\simeq 1.7(\ell_0/h_0)^2$].
The  equation of state is also similar to that of the planar
case, Eq.\ (\ref{3.12}),
\beq
T=T_0-\frac{\eta_0}{4\kappa_0}(u_0-u_z)^2+
\frac{\zeta_T}{\zeta_p}\frac{T_0}{p_0}(p-p_0)+{\cal O}(g^4).
\label{5.4}
\eeq

As for  the momentum and heat
fluxes, the results through third order are
\beq
P_{rr}(r)=p_0\left[1-\frac{92}{25}\frac{\rho_0\eta_0^2g^2}{p_0^3}+\frac{1}{20}
\left(\frac{mg}{k_BT_0}\right)^2r^2\right]+
{\cal O}(g^4),
\label{5.5}
\eeq
\beq
P_{\phi\phi}(r)=p_0\left[1-\frac{92}{25}\frac{\rho_0\eta_0^2g^2}{p_0^3}
+\frac{3}{20}
\left(\frac{mg}{k_BT_0}\right)^2r^2\right]+
{\cal O}(g^4),
\label{5.6}
\eeq
\beq
P_{zz}(r)=p_0\left[1+\frac{184}{25}\frac{\rho_0\eta_0^2g^2}{p_0^3}
+\frac{7}{10}
\left(\frac{mg}{k_BT_0}\right)^2r^2\right]+
{\cal O}(g^4),
\label{5.7}
\eeq
\beq
P_{rz}(r)=\frac{\rho_0g}{2}r\left[1+\frac{\rho_0^2
g^2}{192\eta_0\kappa_0T_0}{r}^4+\frac{2}{25}
\left(\frac{mg}{k_BT_0}\right)^2r^2\right]+
{\cal O}(g^5),
\label{5.8}
\eeq
\beq
q_r(r)=\frac{\rho_0^2g^2}{16\eta_0}r^3+{\cal O}(g^4),
\label{5.9}
\eeq
\beqa
q_z(r)&=&-\frac{2m g
\kappa_0}{5k_B}\left[1-\frac{4(1358-23\alpha)\rho_0\eta_0^2
g^2}{25p_0^3}
-\frac{209-3\alpha}{50}\left(\frac{mg}{k_BT_0}\right)^2r^2
\right.\nonumber\\
&&\left.-\frac{(15-\alpha)\rho_0^2
g^2}{64\eta_0\kappa_0T_0}{r}^4
\right]
+{\cal O}(g^5).
\label{5.10}
\eeqa
As expected,
they strongly differ from the NS results, Eqs.\ (\ref{2.20a}), (\ref{2.21b}),
(\ref{2.26.1}) and (\ref{2.26.2}),
with the exception of $q_r$, in which case the error of the NS value is  of
order $g^4$.

The non-monotonic behavior of $T(r)$ is not only an interesting effect but
also a counterintuitive result, given that the radial component of the heat
flux monotonically increases with the distance from the pipe axis. Consider
the inner cylinder $r\leq r_{\text{max}}$. Within that region the
temperature increases radially and yet the heat flows outwards from the
colder to the hotter points! The solution to this paradox lies in the
dramatic breakdown of the Fourier law (\ref{2.21a}) within the region $r\leq
r_{\text{max}}$. Following Hess and Malek Mansour \cite{HM99}, a heuristic
extension of the Fourier law can be written as
\beq
-\kappa \frac{\partial T}{\partial r}=q_r-\xi^2 \nabla^2 q_r
,
\label{5.11}
\eeq
where $\xi$ is a characteristic distance of the order of the mean free path.
According to Eq.\ (\ref{5.11}), the sign of the thermal gradient results
from the competition between $q_r$ and its Laplacian. The simple estimate
$\nabla^2 q_r=r^{-1}{\partial}\left(r{\partial
q_r}/{\partial r}\right)/{\partial r}\sim q_r/r^2$ shows that $\partial
T/\partial r>0$ for $r< \xi$. It is easy to check that Eq.\ (\ref{5.11}),
with $\xi=\frac{1}{3}r_{\text{max}}\simeq 0.71 \ell_0$,
is indeed consistent with the profiles (\ref{5.3}) and (\ref{5.9}).
If one characterizes the deviation from the Fourier law by means of an {\em
apparent\/} thermal conductivity coefficient defined by
\beq
q_r=-\kappa_{\text{app}} \frac{\partial T}{\partial r},
\label{5.12}
\eeq
then one has
\beq
\frac{\kappa_{\text{app}}}{\kappa}=\left[1-
\left(\frac{r_{\text{max}}}{r}\right)^2\right]^{-1}+{\cal O}(g^2).
\label{5.13}
\eeq
The above ratio vanishes at $r=0$, is negative in the interval
$0<r<r_{\text{max}}$, diverges at $r=r_{\text{max}}$, and finally tends to
unity from above for $r\gg r_{\text{max}}$.

\begin{figure}
\begin{center}
\parbox{0.5\textwidth}{\epsfxsize=\hsize
\epsfbox{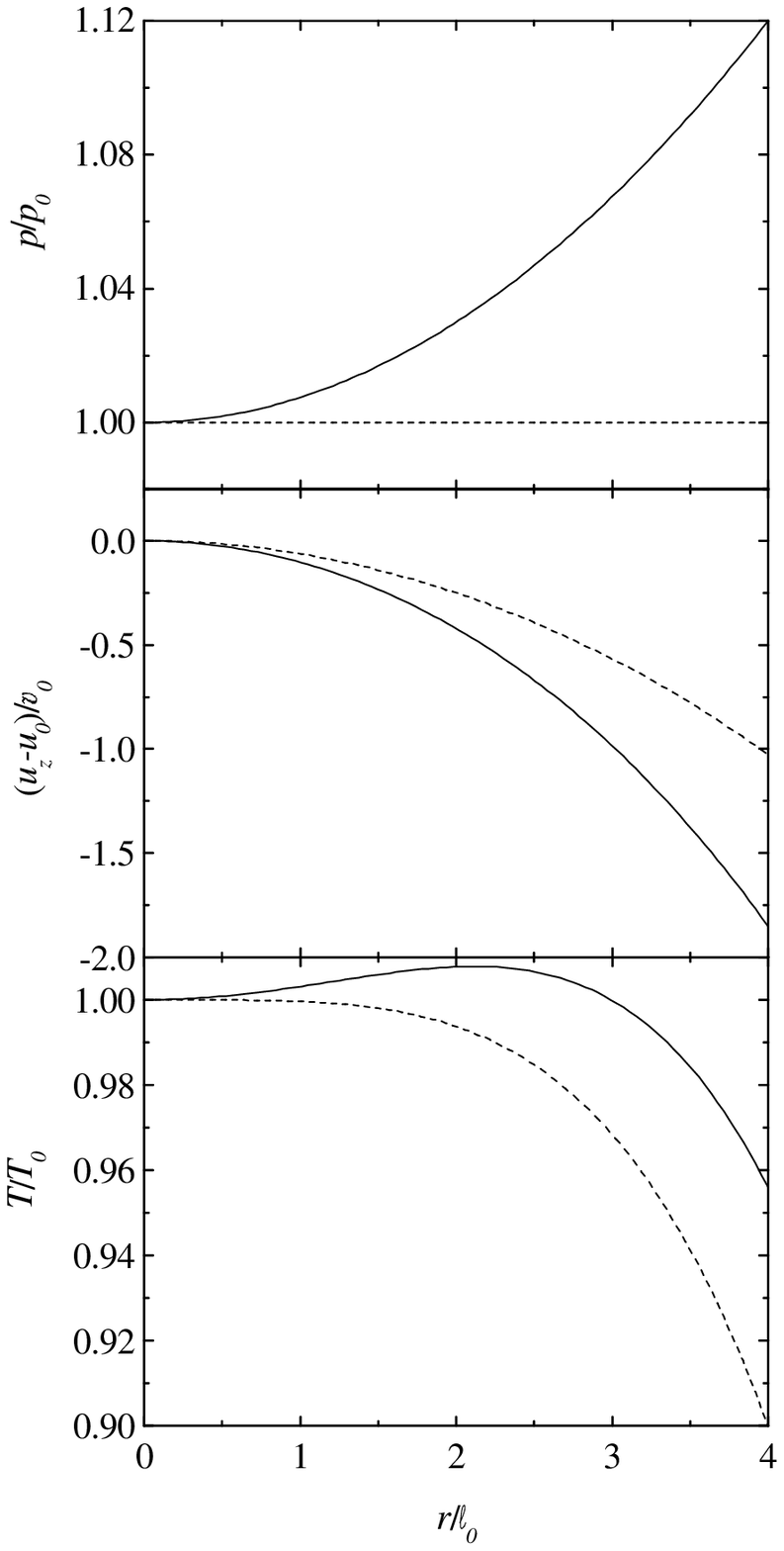}}
\caption{Hydrodynamic profiles for the case $g=0.1\nu_0 v_0$, as predicted
by the Navier-Stokes description (dashed lines) and by the kinetic theory
description (solid lines).
\label{fig1}}
\end{center}
\end{figure}

\begin{figure}
\begin{center}
\parbox{0.5\textwidth}{\epsfxsize=\hsize
\epsfbox{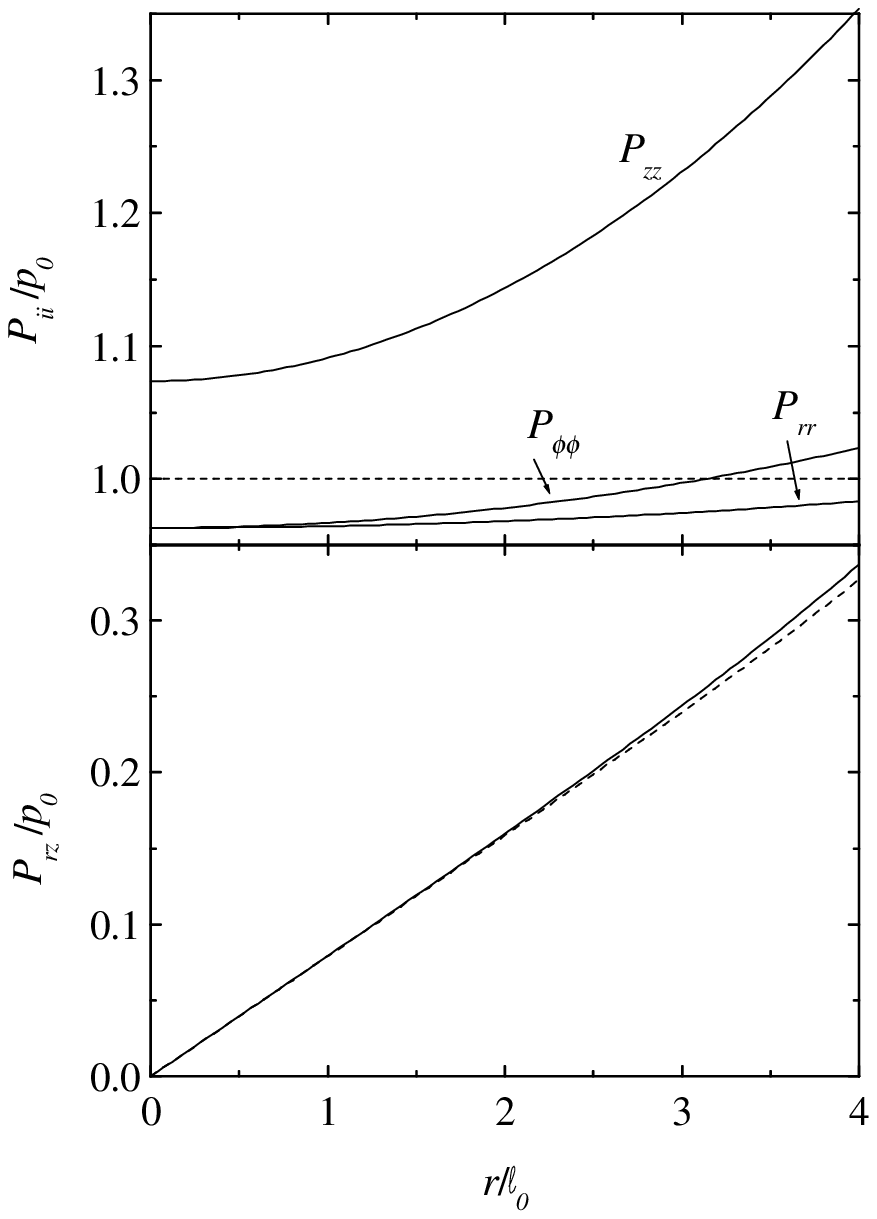}} \caption{Same as in Fig.\ \protect\ref{fig1}
but for the elements of the pressure tensor. \label{fig2}}
\end{center}
\end{figure}

\begin{figure}
\begin{center}
\parbox{0.5\textwidth}{\epsfxsize=\hsize
\epsfbox{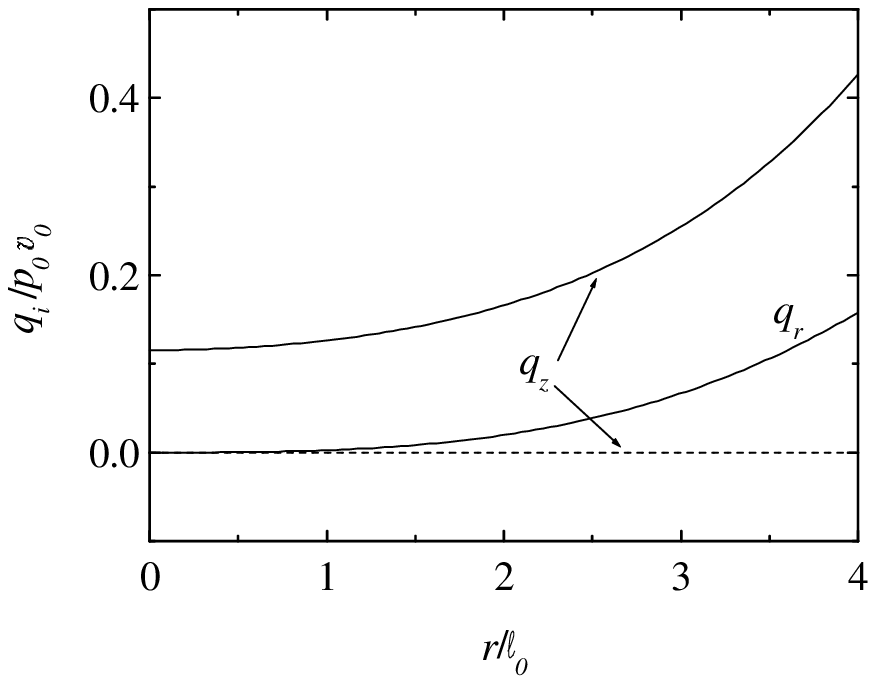}}
\caption{Same as in Fig.\ \protect\ref{fig1} but for the
components of the heat flux.
\label{fig3}}
\end{center}
\end{figure}

The breakdown of the Newton law is characterized by an apparent shear
viscosity coefficient defined by
\beq
P_{rz}=-\eta_{\text{app}}\frac{\partial u_z}{\partial r}.
\label{5.14}
\eeq
The ratio between this coefficients and the NS shear viscosity is
\beqa
\frac{\eta_{\text{app}}}{\eta}&=&1-\zeta_u'\frac{\rho_0^2g^2}{p_0^3}-
\frac{19-\alpha}{20}
\left(\frac{mg}{k_BT_0}\right)^2 r^2+{\cal O}(g^4)\nonumber\\
&=&1-\frac{\zeta_u'}{5\zeta_T}\frac{\ell_0^2}{T_0}\frac{\partial^2
T}{\partial
r^2}-\frac{2(2533-31\alpha)}{175}\frac{\ell_0^2}{v_0^2}\left(\frac{\partial
u_z}{\partial r}\right)^2+{\cal O}(g^4).
\label{5.15}
\eeqa
The last line of Eq.\ (\ref{5.15}) clearly shows that $\eta_{\text{app}}$
incorporates super-Burnett terms. Those terms can be written in equivalent
alternative forms by taking into account that
\beq
\frac{\partial^2 T}{\partial r^2}=-\frac{3\eta_0}{4\kappa_0}\left(
\frac{\partial u_z}{\partial
r}\right)^2+\frac{\zeta_T}{\zeta_p}\frac{T_0}{p_0}\frac{\partial^2 p
}{\partial r^2}+{\cal O}(g^4).
\label{5.16}
\eeq
As an illustration of the corrections over the NS
description provided by
  kinetic theory, we compare  in Figs.\ \ref{fig1}--\ref{fig3} the
hydrodynamic and flux profiles for the case $g=0.1\nu_0(k_BT_0/m)^{1/2}$
[which corresponds to $(h_0/\ell_0)^2=40$], as predicted by both
descriptions when only terms through third order in $g$ are retained.
Although higher order terms are not necessarily negligible for that
particular value of the field, the retained terms can be expected to be
enough, at least at a qualitative level.
The curves for $u_z$ and $q_z$ correspond to hard spheres
($\alpha=\frac{1}{2}$), but they are practically indistinguishable from
those of Maxwell molecules ($\alpha=0$). Except for the shear stress
$P_{rz}$ and the radial heat flux $q_r$, the kinetic theory predictions
dramatically differ from the NS ones. The hydrostatic pressure grows
quadratically rather than being uniform, the flow velocity decreases more
rapidly than expected from the NS description, and the temperature exhibits
a non-monotonic behavior. Normal stress differences appear, the
flux of longitudinal momentum along the longitudinal direction being larger
than the other two normal stresses ($P_{zz}>P_{\phi\phi}>P_{rr}$). Finally, 
the
longitudinal heat flux is not only different from zero (despite the absence
of longitudinal gradients), but it can be even larger than the radial heat
flux.
These features are qualitatively similar to those found in the rectangular
geometry (cf.\ Sec.\ \ref{sec3}), which have been confirmed by computer
simulations \cite{MBG97,RC98,UG99}.

\section{Conclusions}
\label{sec6}
In this paper we have solved the BGK kinetic equation for the (laminar)
steady {\em cylindrical\/} Poiseuille flow fed by a constant gravity field.
The solution has been obtained as a perturbation expansion in powers of the
field through fourth order and for a general class of repulsive potentials.
The results exhibit a very weak sensitivity to the interaction potential and
strongly indicate that the expansion is only asymptotic. A comparison with
the profiles obtained from the Navier-Stokes (NS) constitutive equations
shows that the latter widely fail over distances comparable to the mean free
path. At a qualitative level, the most important limitation of the NS
description is that it predicts a monotonically decreasing temperature as
one moves apart from the cylinder axis. In contrast, the kinetic theory
description shows that the temperature has a local minimum ($T=T_0$) at the
axis ($r=0$) and reaches a maximum value ($T=T_{\text{max}}$) at a distance
 from the center  ($r=r_{\text{max}}$) of the order of the mean free path. In
the region $r\leq r_{\text{max}}$, the radial heat flows from the colder to
the hotter points, what dramatically illustrates the breakdown of the
Fourier law.
Furthermore, a longitudinal component of the heat flux exists in the absence
of gradients along the longitudinal direction.
Non-Newtonian effects are exemplified by the non-uniformity of
the hydrostatic pressure and by the presence of normal stress
differences.

The above effects are similar to those previously found in the case of a
rectangular channel.
This is a non-trivial result, since both geometries are quite different, as
can be expected from the different mathematical structure of the balance
equations [Eqs.\ (\ref{2.1b})--(\ref{2.2}) versus Eqs.\
(\ref{2.15a})--(\ref{2.16})] and of the kinetic equations [Eq.\ (\ref{3.8})
versus Eq.\ (\ref{4.1})]. In the rectangular geometry the relevant space
variable $y$ takes both positive and negative values, while the radial
variable $r$ is positive definite. Also, the normal stress along the
gradient direction is uniform in the rectangular case
($P_{yy}=\mbox{const}$) and non-uniform in the cylindrical case
($P_{rr}\neq\mbox{const}$).
At a quantitative level, on the other hand, the deviations of the NS
profiles from the kinetic theory ones are weaker in the cylindrical geometry
than in the rectangular geometry. For instance, the relative difference
$(T_{\text{max}}-T_0)/T_0$ is about 5 times smaller in the former case than
in the latter.

The analysis carried out in this work can be extended to the Boltzmann
equation for Maxwell molecules, as already done in the channel case
\cite{TSS98}. The hydrodynamic profiles will still be given by Eqs.\
(\ref{5.1})--(\ref{5.3}), but with different numerical values for the
coefficients $\zeta_p$, $\zeta_u$, $\zeta_u'$ and $\zeta_T$.
For hard spheres, the solution can be obtained by approximate schemes, such 
as the moment method \cite{HM99}. Finally,
we hope that the results reported in this paper may stimulate the
undertaking of computer simulations of the Poiseuille flow induced by
gravity in a pipe.
\ack
This work has been done under the auspices of the Agencia Espa\~nola de
Cooperaci\'on Internacional (Programa de Cooperaci\'on Interuniversitaria
Hispano-Marroqu\'{\i}).
A.S. acknowledges partial support from the DGES (Spain) through
Grant No.\ PB97-1501 and from the Junta de Extremadura (Fondo Social Europeo)
through Grant No.\ IPR99C031.

\end{document}